\documentclass[aps,twocolumn,showpacs]{revtex4}
\usepackage{graphicx}
\usepackage{epstopdf}
\usepackage{epsfig}
\usepackage[english]{babel}

\begin{document}

\def\bsigma{\mbox{\boldmath $\sigma$}}
\def\bomega{\mbox{\boldmath $\omega$}}
\def\ss{\scriptscriptstyle }






\title{Spatial dispersion of magnetic-edge magnetoplasmons: Effect of semi-infinite gate }
\author{O. G. Balev}
\email[Electronic address:]{ogbalev@ufam.edu.br}
\affiliation{Departamento de F\'{\i}sica, Universidade Federal do Amazonas, 69077-000,
Manaus, Amazonas, Brazil}
\author{I. A. Larkin$^{2,3}$}
\affiliation{$^{2}$Department of Physics, Minho University, 4710-057, Braga, Portugal} 
\affiliation{$^{3}$ Institute of Microelectronics Technology Russian Academy of Sciences, 142432 Chernogolovka, Russia}

\begin{abstract} 
Magnetic-edge magnetoplasmons (MEMPs) are obtained for a two-dimensional electron system (2DES)
with atop semi-infinite metallic gate, at a distance $d$,  and atop semi-infinite ferromagnetic film
 at a  strong perpendicular magnetic field. 
For two most fast MEMPs, one with positive chirality and other with negative chirality, a strong spatial dispersion, due to
effect of metallic half-plane gate, is obtained; some slower MEMPs manifest spatial dispersion too.
Present MEMPs are localized at the magnetic-edge that is  close to the wedge of metallic half-plane gate;
the metallic wedge enhances localization of MEMPs at magnetic-edge.
Obtained spatial dispersion has unconventional form. In particular, for two most fast  MEMPs the phase 
velocities, $\omega/k_{x}$, are the linear polynomials on the wave vector $k_{x}$
in the long-wavelength region, $k_{x}d \ll 1$.
Strong effect of the ferromagnetic film hysteresis on the MEMPs phase velocities and their anti-crossings are obtained
for $0<k_{x}d \leq1$.
Two MEMPs of opposite chirality, especially two most fast MEMPs, at some resonance frequency can
create a resonance circuit, with closed wave path along a fraction of the magnetic edge perimeter,  
with a total change of the wave phase given by an integer of $2\pi$.
\end{abstract}

\date{\today}
\maketitle


\section{Introduction}

Different types of "density gradient" edge magnetoplasmons (EMPs) have been
studied for (conventional) 2DES \cite{volkov88,volkov91,kushwaha2001,ashoori92,aleiner94,zhitenev95,ernst1996,balev97,bal,balev99,haug2004,kukush09,%
bal2010,kukush2012a,kukush2012b,feve2013,sasaki2016}
subjected to a strong homogeneous magnetic field; for a discussion of previous publications, e.g., see \cite{volkov88}. 
These EMPs appear due to a
strong change of the stationary local electron density at the edge of the
channel (in particular, in the vicinity of edge states) that induces a
strong modulation of the local magnetoconductivity tensor; modulation of the
nondiagonal components is especially important \cite{volkov88,volkov91,aleiner94,balev97,bal,bal2010}. 
In addition, EMPs in graphene have attracted attention recently
\cite{balev2011,balev2012a,balev2012b,apell2012a,apell2012b,glattli2013,kumada2013,glattli2014}. 

As a rule, EMPs are studied at a long-wavelength region $k_{x} a_{EM} \ll 1$, 
\cite{volkov88,ashoori92,aleiner94,zhitenev95,ernst1996,balev97,bal,balev99,bal2010}
 where $a_{EM}$ is a finite characteristic scale; it is understood that another characteristic 
 scale $1/k_{x}$, related with the wavelength $2\pi/k_{x}$, also is involved. 
 Typically $a_{EM}$ is the characteristic localisation, along $y$,  for a charge density
of the most fast EMP; for different models of EMPs  the value of $a_{EM}$ can be quite different as well as the physical
picture involved  \cite{volkov88,aleiner94,balev97,bal,balev99,bal2010}.  
For a homogeneous sample without the gate,  in the long-wavelength region, only the fundamental EMP (the only EMP in the
 model of Ref. \cite{volkov88}; notice, this EMP in Ref. \cite{aleiner94}
 is called as \textit{the conventional edge magnetoplasmon mode}) shows spatial dispersion as
a weak logarithmic dependence of the phase velocity \cite{volkov88,volkov91,kushwaha2001,aleiner94,balev97,bal,bal2010}  
on  $k_{x}$, $\omega/ k_{x} \propto \ln(1/k_{x} a_{EM} )$. 
In addition, here the phase velocities
of the rest of EMPs  are independent of $k_{x}$ and smaller \cite{aleiner94,balev97,bal,bal2010}. Often, these 
dispersionless modes  are called as acoustic EMPs \cite{aleiner94}.
If a plane metallic gate is present at a distance $d$ from the 2DES then all EMPs are
dispersionless for $k_{x} d \ll 1$, as their phase velocities are independent of $k_{x}$ 
\cite{zhitenev95,balev97,bal,bal2010}.  

Recently,\cite{balev2013,balev2014} the chiral modes in the homogeneous 2DES (localized at $z=0$ plane in GaAs based sample, with 
a very thing homogeneous nonmagnetic metallic gate above it, at $z=d$)
 induced by "magnetic gradient" have been  obtained theoretically. Laterally inhomogeneous strong magnetic field within the
 plane of 2DES appears due to ferromagnetic semi-infinite film ($y \leq 0$) of a finite thickness,  located atop of the gate.
  Named as magnetic-edge magnetoplasmons (MEMPs), 
these modes are localized,  along the $y-$direction, on the characteristic scale $d$ in a vicinity of $y=0$; i.e., 
in the principal region of magnetic field inhomogeneity \cite{balev2013,balev2014}. 
For $k_{x} d \ll 1$ all these MEMPs show acoustic dispersion  \cite{balev2013,balev2014}.
I.e., for a plane metallic gate all MEMPs are dispersionless in the long-wavelength region.
Notice, any MEMPs are absent if a strong magnetic field applied to 2DES
is spatially homogeneous  \cite{balev2013,balev2014}.

In present study we consider the effect on MEMPs by a half-plane metallic gate (HPMG), at  ($y<0$, $z=d$). 
Schematic view of the model geometry, otherwise similar to the one of  Refs.  \cite{balev2013,balev2014}, is shown in
Fig. 1. Here 2DES, at $z=0$, is embedded in 
GaAs based sample, with the dielectric constant $\varepsilon$,  that occupies a half-space $z<d$.
As the semi-infinite gate, i.e., HPMG, is assumed very thin it is not shown explicitly in Fig. 1. 
 2DES is subjected to a strong
laterally inhomogeneous magnetic field $\mathbf{B}(y)=B(y)\mathbf{\hat{z}}$,
which appears due to ferromagnetic semi-infinite film of a finite thickness $\eta d$. 
A strong external spatially homogeneous magnetic field $\mathbf{B}_{ext}=B_{ext}\mathbf{\hat{z}}$ is applied as well.

Qualitative and strong
quantitative effects of HPMG on some MEMPs are obtained. In particular, a strong spatial dispersion  of unconventional form,
for two MEMPs of opposite chirality, is obtained for $k_{x} d \ll 1$. 

\begin{figure}[ht]
\vspace*{-0.3cm}
\includegraphics [height=4cm, width=8cm]{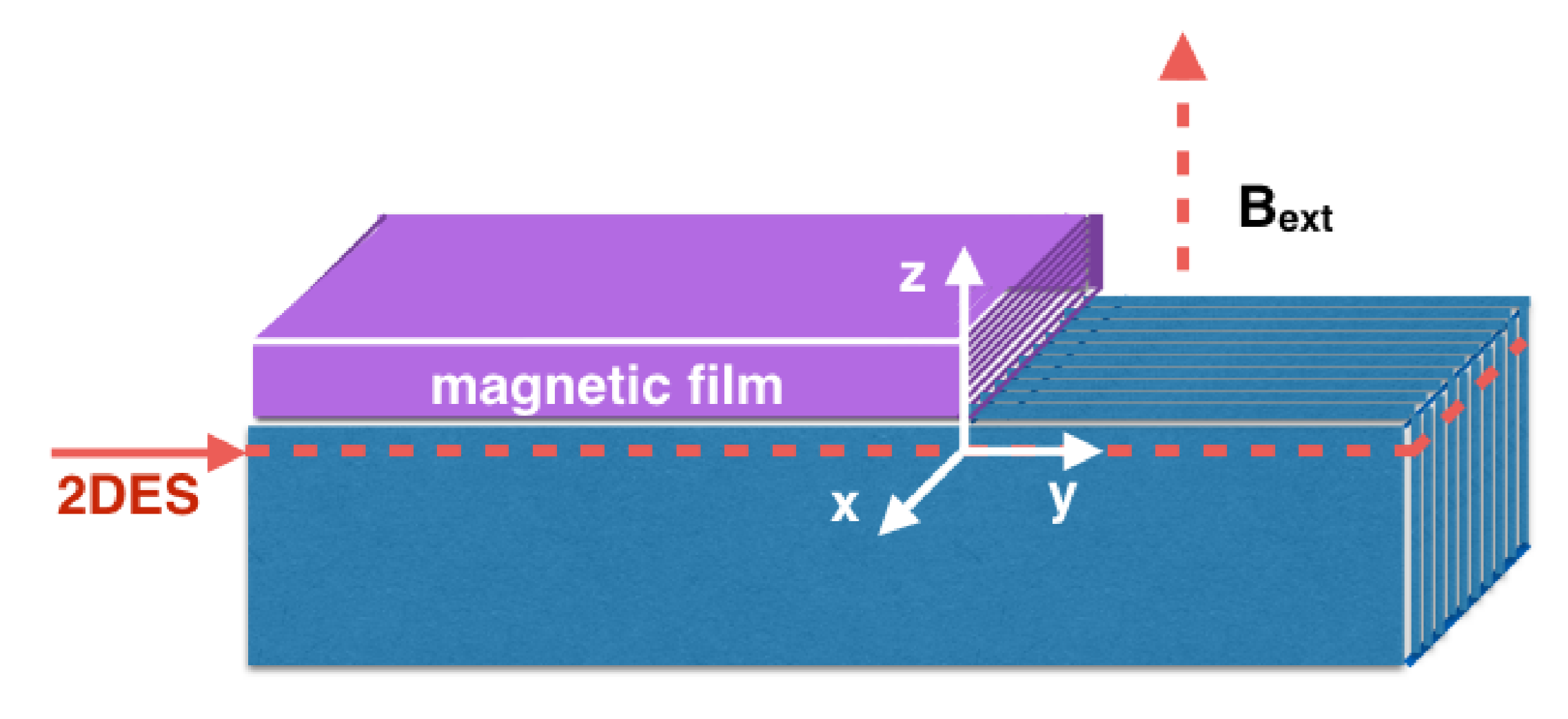}
\vspace*{-0.5cm}
\caption{(Color online) Sketch of model geometry: 2DES, at $z=0$, is embedded in a dielectric
medium.
The ferromagnetic semi-infinite film 
is located at  $y <0$, $d<z<d(1+\eta)$; HPMG, at $y<0$, $z=d$, is not shown. }
\end{figure}

In Sec. II we present further attributes of the model for MEMPs  in 2DES with half-plane atop metallic gate
and then the advanced attributes of the model and pertinent basic relations.
Then in Sec. III we present results for spatial dispersion of MEMPs, effect of hysteresis, anti-crossings. 
In Sec. IV we make concluding remarks.

\section{ Magnetic-edge magnetoplasmons for half-plane gate: Formalism}

\subsection{Model attributes: Introduction}

It is assumed that for any actual spatially homogeneous (in particular, zero) effective magnetic field $B$ applied to 2DES,
we have spatially homogeneous 2DES of a constant area density $n_{I}$
as well as the ions jellium background with the same area density, $n_{I}$.  
Spatially inhomogeneous effective magnetic field $B(y)$ is assumed as a smooth function of $y$, with the characteristic scale
$\Delta y $. Comparing the perturbation of the electron density 
due to spatially inhomogeneous magnetic field from its unperturbed value $n_{I}$, it follows that
the density of 2DES is very weakly modified if the Bohr radius $a_{B} \ll 2\Delta y$ and the quantum magnetic length
$\ell_{0}(y) \ll 2 \Delta y$; $a_{B}=\hbar^{2}/m^{\ast}e^{2}$, $\ell_{0}(y)=\sqrt{\hbar c/|e| B(y)}$, where
$m^{\ast}$ is an effective mass of 2D electrons.
Below we assume that these conditions are satisfied and, respectively, the static electron density is taken 
as $n_{I}$.  

We assume, cf. with Fig. 1, a constant magnetic moment for the ferromagnetic semi-infinite film
$\mathbf{M}_{0}=M_{0} \mathbf{\hat{z}}$; $M_{0}>0$ if otherwise is not stated. Then readily it follows \cite{balev2013,jackson1999} that
\begin{equation}
B(y)=B_{ext}-2 M_{0}\{\arctan(Y)-\arctan(Y/(1+\eta))  \} ,
\label{eq1}
\end{equation}%
where $Y=y/d$; i.e., in present study $\Delta y =d$.  
Point out that in Eq. (\ref{eq1}) in the SI units $M_{0}$  is changed on $\mu_{0} M_{0}/(4\pi)$. 
Notice, in experiments  with GaAs-based 2DES  the values of $d$ and $\eta d$ can be  $\sim 100$nm \cite{Lloyd2012}.
External spatially homogeneous magnetic field $B_{ext}>0$, if otherwise is not stated, is applied as well.


We assume that the low-frequency, 
$\omega \ll \omega _{c}$, and the long-wavelength, 
$k_{x} \ell_{0}(y) \ll 1$, conditions are satisfied. Then the wave 
current density in the quasi-static 
approximation (e.g., cf. with Ref. \cite{balev97,bal,balev00}) is given as 
\begin{eqnarray}
&&j_{x}(y)=\sigma _{xx}(y)E_{x}(y)-\sigma _{yx}^{0}(y)E_{y}(y) , \nonumber \\
&&j_{y}(y)=\sigma _{yy}(y)E_{y}(y)+\sigma _{yx}^{0}(y)E_{x}(y),  
\label{eq2}
\end{eqnarray}
where we have suppressed 
the exponential factor $\exp [-i(\omega t-k_{x}x)]$;
arguments $\omega$, $k_{x}$ in $j_{\mu}(y)$, $E_{\mu}(y)$ are omitted, to simplify notations.
We will neglect by a dissipation assuming a clean 2DES and sufficiently low
temperatures $T$.

In the absence of metallic gates
and for homogeneous background dielectric constant $\varepsilon$,
using Eq. (\ref{eq1}), the Poisson equation and the linearized continuity equation, 
we obtain for the wave charge density $\rho (\omega ,k_{x},y)$ the integral (cf. with Ref. \cite{balev97,bal,balev00}) equation
\begin{eqnarray}
&&-i\omega \rho (\omega ,k_{x},y)-i\frac{2}{\varepsilon }
\{k_{x}\frac{d}{dy}[\sigma _{yx}^{0}(y)] \} \nonumber \\
&& \times \int_{-\infty }^{\infty }dy^{\prime }K_{0}(|k_{x}||y-y^{\prime }|)
\rho(\omega ,k_{x},y^{\prime })=0,  
\label{eq4}
\end{eqnarray}
where $G^{(0)}(k_{x},y-y^{\prime},z=0;z^{\prime}=0)=(2/\varepsilon)K_{0}(|k_{x}||y-y^{\prime }|)$ is 
the Fourier transform over the coordinate $\Xi=x-x^{\prime}$ of the 
Green function at $(x,y,z=0)$ for the unit charge localized at $(x^{\prime},y^{\prime},z^{\prime}=0)$;
$K_{0}(x)$ is the modified Bessel function. 
In the model of  Refs.  \cite{balev2013,balev2014} is assumed that in addition to the semi-infinite magnetic film, a very thin infinite metallic 
nonmagnetic film of the thickness $\ll d$ is places on the top of the sample
at a distance $d$ from the 2DES. Then the kernel $K_{0}$ in Eq. (\ref{eq4}) is replaced by 
\begin{eqnarray}
R_{g}^{(1)}(|y-y^{\prime}|&,&k_{x};d)=K_{0}(|k_{x}||y-y^{\prime }|) \nonumber \\
&&- K_{0}(|k_{x}|\sqrt{(y-y^{\prime })^{2}+4d^{2}})  ,  
\label{eq5}
\end{eqnarray}
here the Green function $G^{(1)}(k_{x},y-y^{\prime},z=0;z^{\prime}=0)=(2/\varepsilon)R_{g}^{(1)}(|y-y^{\prime}|,k_{x};d)$ .

In present model  HPMG is a half-plane metallic nonmagnetic film of negligible thickness. 
Besides HPMG and the plane of 2DES the rest of space presents dielectric background with  
spatially homogeneous dielectric constant $\varepsilon$. Henceforth
the ferromagnetic semi-infinite film is taken as a dielectric too,  with dielectric constant  $\varepsilon$.

\subsection{Basic Relations and Advanced Attributes of the Model }

For the present model, with HPMG,  we obtain the integral equation for MEMPs, cf. with Eq. (14) of Ref. \cite{balev2014},
as
\begin{eqnarray}
&&W \rho (\omega ,k_{x},X)-g_{0}(X)f_{0}(X)    \int_{-1 }^{1 } dX^{\prime }   \nonumber \\
&& \times \widetilde{G}^{(2)}(k_{x};d\tan(\frac{\pi}{2}X),z=0;d\tan(\frac{\pi}{2}X^{\prime}),z^{\prime}=0)   \nonumber  \\
&& \times  \frac{(\pi/2)}{\cos^ {2}(\pi X^{\prime }/2)} \rho (\omega,k_{x},X^{\prime })=0,  
\label{eq6}
\end{eqnarray}%
where 
$W=\omega/(k_{x}v_{0})$ is the dimensionless phase velocity and its sign corresponds 
to the chirality of a wave, i.e., $W>0$ ($W<0$) for positive (negative) chirality if $v_{0} >0$.
Here $v_{0}=|e|cn_{I}/(\varepsilon B_{0})$ is a characteristic velocity of the problem with  
$B_{0}=B_{ext}^{2}/(4M_{0})$.
Further,   $X=\frac{2}{\pi} \arctan(Y)$ [$X^{\prime}=\frac{2}{\pi} \arctan(Y^{\prime})$] is a  new variable
that change from $-1$ to $1$ as $Y$ [$Y^{\prime}$] changes from $-\infty$ to $\infty$, 
\begin{equation}
g_{0}(X)=\{\frac{1+\eta }{(1+\eta )^{2}+\tan^{2}(\pi X/2)}-\frac{1}{1+\tan^{2}(\pi X/2)}\}
\label{eq7}
\end{equation}%
is a dimensionless gradient of $B(y)$, and the factor%
\begin{equation}
f_{0}(X)=\{1-\frac{2M_{0}}{B_{ext}}[\frac{\pi}{2} X-\arctan (\frac{1}{1+\eta } \tan(\frac{\pi}{2}X) )]\}^{-2}.
\label{eq8}
\end{equation}
Point out that $g_{0}(X)$ is the symmetric function of $X$ as function $f_{0}(X)$ is neither symmetric nor antisymmetric.
Here $\widetilde{G}^{(2)}(k_{x};y,z=0;y^{\prime},z^{\prime}=0) \equiv (\varepsilon/2)G^{(2)}(k_{x};y,z=0;y^{\prime},z^{\prime}=0)$,
where $G^{(2)}(k_{x};y,z=0;y^{\prime},z^{\prime}=0)$ is 
the Fourier transform, over the coordinate $\Xi=x-x^{\prime}$, of the 
Green function at $z=0$ plane (of 2DES) for the unit charge localized at $(x^{\prime},y^{\prime},z^{\prime}=0)$. 
In what follows we will use notations $\Delta_{\varphi}(k_{x},X,X^{\prime}) \equiv \widetilde{G}^{(2)}(k_{x};d\tan(\frac{\pi}{2}X),z=0;d\tan(\frac{\pi}{2}X^{\prime}),z^{\prime}=0)$, and 
\begin{equation}
\Delta_{\varphi}(k_{x},X,X^{\prime})=\int_{0}^{\infty} d \xi \cos(k_{x}d \; \xi) \widetilde{\Delta}_{\varphi}(\xi,X,X^{\prime}) ,
\label{eq9}
\end{equation}
where dimensionless variable $\xi=\Xi/d$ is introduced. 
Using the analytical solution for the electric potential potential of the point charge nearby the wedge
of conducting half-plane in a vacuum, e.g., see \cite{landau1960}, we straightforwardly obtain that

\begin{figure}[ht]
\vspace*{-0.3cm}
\includegraphics [height=14cm, width=10cm]{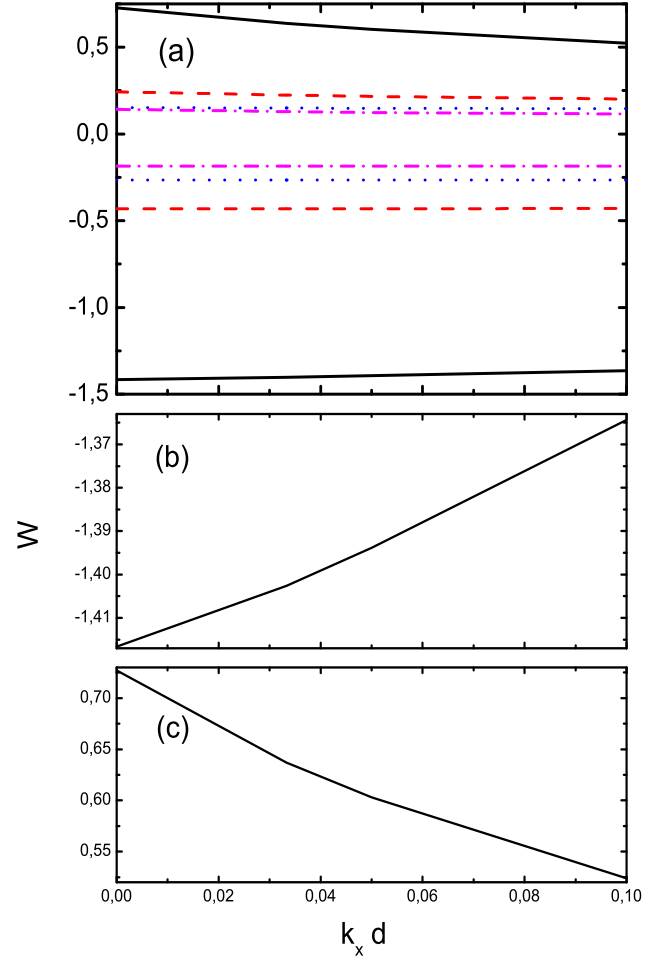}
\vspace*{-1.0cm}
\caption{(Color online) Dimensionless phase velocity $W=\omega/k_{x} v_{0}$ as function of $k_{x}d$ 
at a long-wavelength region $0.1 \geq k_{x} d > 0$ for
$\eta=1$, $2M_{0}/B_{ext}=0.5$, $B_{ext}=0.6$T, $v_{0}>0$. Figs. 2(a), 2(b), and 2(c) show: 
(a) four fastest MEMPs among the modes with the positive (negative) chirality,
(b) zoom for the fastest MEMP with the negative chirality, and (c) zoom for the fastest MEMP with the positive chirality.}
\end{figure}

\begin{figure}[ht]
\vspace*{-0.3cm}
\includegraphics [height=14cm, width=10cm]{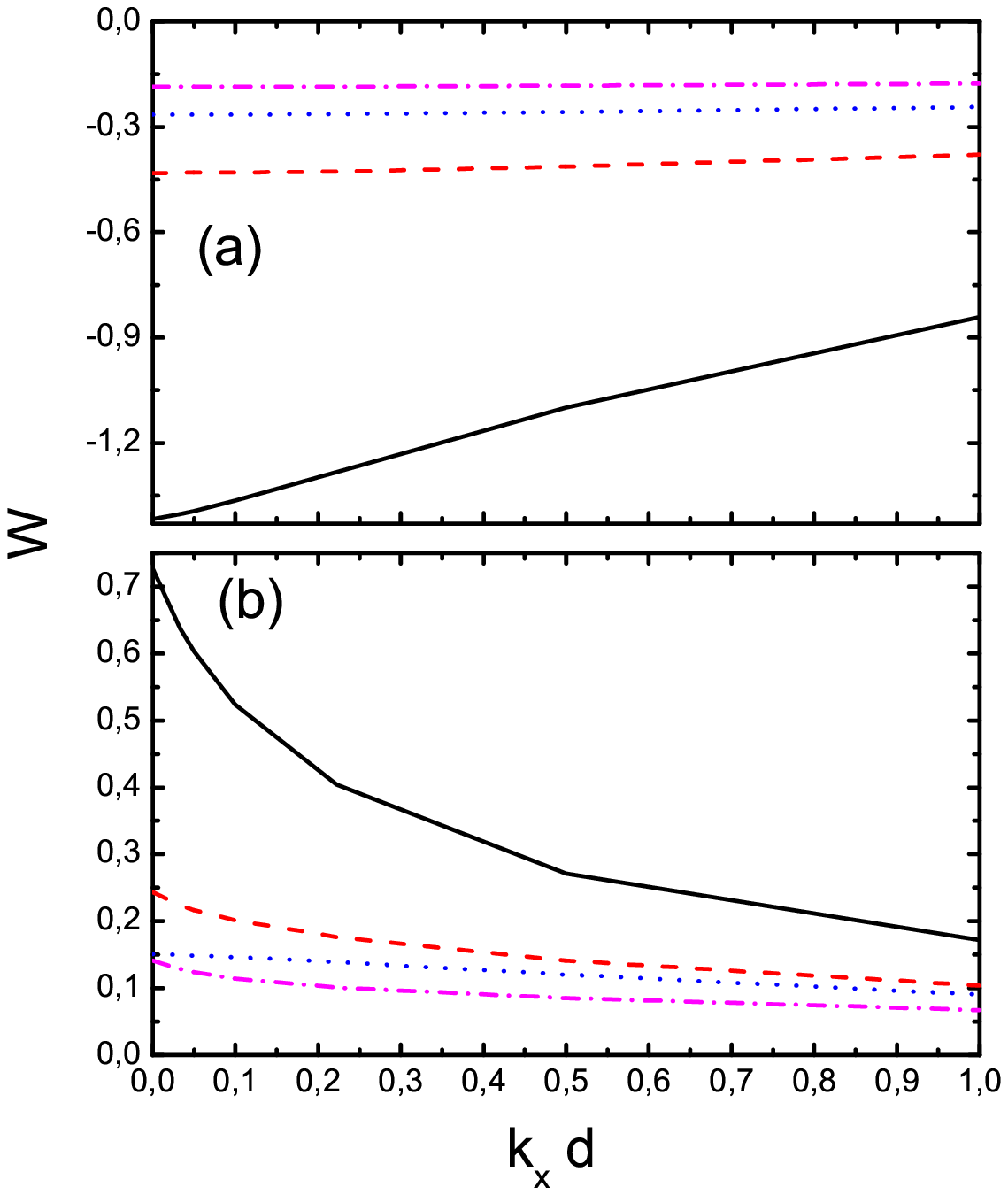}
\vspace*{-3.0cm}
\caption{(Color online) Dimensionless phase velocity $W$ as function of $k_{x}d$ at a wide region
$1.0 \geq k_{x} d > 0$ for parameters of Fig. 2.
Figs. 3(a) and 3(b) show 
four fastest MEMPs among the modes with the negative  and the positive chirality, respectively.}
\end{figure}

\begin{eqnarray}
&&\widetilde{\Delta}_{\varphi}(\xi,X,X^{\prime})=\frac{1}{\pi} \left\{[\xi^{2}+(\tan(\frac{\pi}{2}X^{\prime})-\tan(\frac{\pi}{2}X))^{2} ]^{-1/2} \right. \nonumber \\
&& \times \arccos[-\frac{\cos(\pi(X^{\prime}-X)/4)}{\cosh(\chi/2)}]  \nonumber \\
&&-\left[\xi^{2}+4+(\tan(\frac{\pi}{2}X^{\prime})-\tan(\frac{\pi}{2}X))^{2} \right]^{-1/2} \nonumber \\
&& \left. \times \arccos[\frac{\sin(\pi(X^{\prime}+X)/4)}{\cosh(\chi/2)}] \right\}    ,  
\label{eq10}
\end{eqnarray}%
where 
\begin{eqnarray}
e^{\chi}&=&  \left\{\left[  \frac{\xi^{2}+2+\tan^{2}(\frac{\pi}{2}X^{\prime})+\tan^{2}(\frac{\pi}{2}X)  }{2 \sqrt{1+\tan^{2}(\frac{\pi}{2}X^{\prime})}
\sqrt{1+\tan^{2}(\frac{\pi}{2}X)}}  \right]^{2} -1 \right\}^{1/2}  \nonumber \\
&&+ \frac{\xi^{2}+2+\tan^{2}(\frac{\pi}{2}X^{\prime})+\tan^{2}(\frac{\pi}{2}X)  }{2 \sqrt{1+\tan^{2}(\frac{\pi}{2}X^{\prime})}
\sqrt{1+\tan^{2}(\frac{\pi}{2}X)}}  \;\;\;  . 
\label{eq11}
\end{eqnarray}%

For $k_{x} \to 0$ we also obtain $\Delta_{\varphi}(k_{x}=0 ,X,X^{\prime})$ in the form different from
the integral form of Eq. (\ref{eq9}). It is given as 
\begin{eqnarray}
&&\Delta_{\varphi}(k_{x}=0,X,X^{\prime})= - \Re \left\{   \ln  \right. \nonumber \\
&&
 \left.  \left[\frac{(\tan(\frac{\pi}{2}X)-i)^{1/2} - (\tan(\frac{\pi}{2}X^{\prime})-i)^{1/2} }{(\tan(\frac{\pi}{2}X)-i)^{1/2} + %
(\tan(\frac{\pi}{2}X^{\prime})+i)^{1/2}}  \right] \right\} =  \label{eq12} \\
&& -\frac{1}{2} \ln  \left\{ \frac{(Z_{+}(X)-Z_{+}(X^{\prime}))^{2}+(Z_{-}(X)-Z_{-}(X^{\prime}))^{2}}{(Z_{+}(X)+Z_{+}(X^{\prime}))^{2}+(Z_{-}(X)-Z_{-}(X^{\prime}))^{2}}       \right\}   , \nonumber 
\end{eqnarray}
where $Z_{\pm}(t)=[2\sqrt{\tan^{2}(\pi t/2)+1} \pm 2\tan(\pi t/2)]^{1/2}$. Point out,  Eq. (\ref{eq12}) is equivalent
to  Eq. (\ref{eq9}) for $k_{x} = 0$.

From Eqs. (\ref{eq6})-(\ref{eq10}) it follows that  a solution of the integral equation Eq. (\ref{eq6}) is not either symmetric
or antisymmetric.  We look for  a solution of Eq. (\ref{eq6}) in the form
\begin{eqnarray}
&&\rho(\omega ,k_{x};X)=g_{0}(X)f_{0}(X)\{\sum_{n=0}^{N_{0} }a_{n}(\omega,k_{x}) 
\left[ \cos (n \pi X) \right. \nonumber \\
&& \left.   -\frac{1}{2}\delta _{n, 0}\right]+\sum_{n=N_{0}+1}^{2N_{0}+1 }a_{n}(\omega ,k_{x}) \sin \left[ (n-N_{0})\pi X \right] \} ,
\label{eq13}
\end{eqnarray}%
which for $N_{0} \to \infty$ is  exact as pertinent basis of the orthonormal functions  becomes complete. 

For a finite $N_{0}$, Eq. (\ref{eq13}) is an expansion
 over the set of $(N_{0}+1)$ symmetric and $(N_{0}+1)$ antisymmetric  orthonormal functions. 
Now Eq. (\ref{eq6}) we reduce to the system of $2(N_{0}+1)$ linear homogeneous
equations.  The first $(N_{0}+1)$ equations, for $m=0, 1, 2,...,N_{0}$, are obtained by multiplying Eq. (\ref{eq6}) on
$ \cos (m\pi X)/[g_{0}(X)f_{0}(X)]$ and then integrating over $X$, $\int_{-1}^{1} dX...$, as follows
\begin{equation}
Wa_{m}-\{\sum_{n=0}^{N_{0} } r_{m,n} a_{n}+\sum_{n=N_{0}+1}^{2N_{0}+1} r_{m,n} a_{n}\}=0 , 
 \label{eq14}
\end{equation}%
The other  $(N_{0}+1)$ equations, for $m=N_{0}+1, N_{0}+2,...,2N_{0}+1$, are obtained after multiplying Eq. (\ref{eq6}) by
$\sin \left[ (m-N_{0})\pi X \right]/(g_{0}(X)f_{0}(X))$ and then integrating over $X$. As a result we obtain the system of equations that formally 
can be obtained from Eq. (\ref{eq14}) by pertinent change in the values of $m$. Finally, the system 
of $2(N_{0}+1)$ linear homogeneous equations is given as
\begin{equation}
Wa_{m}-\sum_{n=0}^{2N_{0}+1} r_{m,n} a_{n}=0 ,
 \label{eq15}
\end{equation}%
where $m,n=0, 1, 2, 3,...,2N_{0}+1$. The matrix elements $r_{mn}=I^{(1)}_{mn}$ for ($m \leq N_{0}$, $n \leq N_{0}$), 
$r_{mn}=I^{(2)}_{mn}$ for ($m \geq N_{0}+1$, $n \leq N_{0}$),
$r_{mn}=I^{(3)}_{mn}$ for ($m \leq N_{0}$, $n \geq N_{0}+1$), 
$r_{mn}=I^{(4)}_{mn}$ for ($m \geq N_{0}+1$, $n \geq N_{0}+1$) are given by Eqs. (\ref{eqa1})-(\ref{eqa4}) 
in Appendix \ref{app1}, where, in particular, an integral over $X$ ($X^{\prime}$) 
from $-1$ to $1$ is transformed to the interval from $0$ to $1$.

For a nontrivial solution Eq. (\ref{eq13}) the determinant of the $2(N_{0}+1)\times 2(N_{0}+1)$ matrix  
of Eq. (\ref{eq15}) must be equal to
zero. This gives  $2(N_{0}+1)$  dispersion relations for dimensionless phase velocities $W=W_{j}(k_{x})$, $j=1,2,..., 2(N_{0}+1)$ 
of the first $2(N_{0}+1)$ MEMPs branches. These dispersion relations readily can be rewritten in a more usual 
form $\omega=\omega_{j}(k_{x})$.
Then, e.g., for $N_{0}=1$ the first four MEMPs  are obtained and for $N_{0}=2$ we will obtain the first six MEMPs, however, only
two of them are new modes as other four solutions give previously obtained four MEMPs with better precision. 
With further increase in $N_{0}$ we will calculate, in particular, these first four MEMPs  too, noteworthy, with fast growing precision.

\section{Spatial dispersion of magnetic-edge magnetoplasmons. Effect of a hysteresis }

\begin{figure}[ht]
\vspace*{-0.3cm}
\includegraphics [height=14cm, width=10cm]{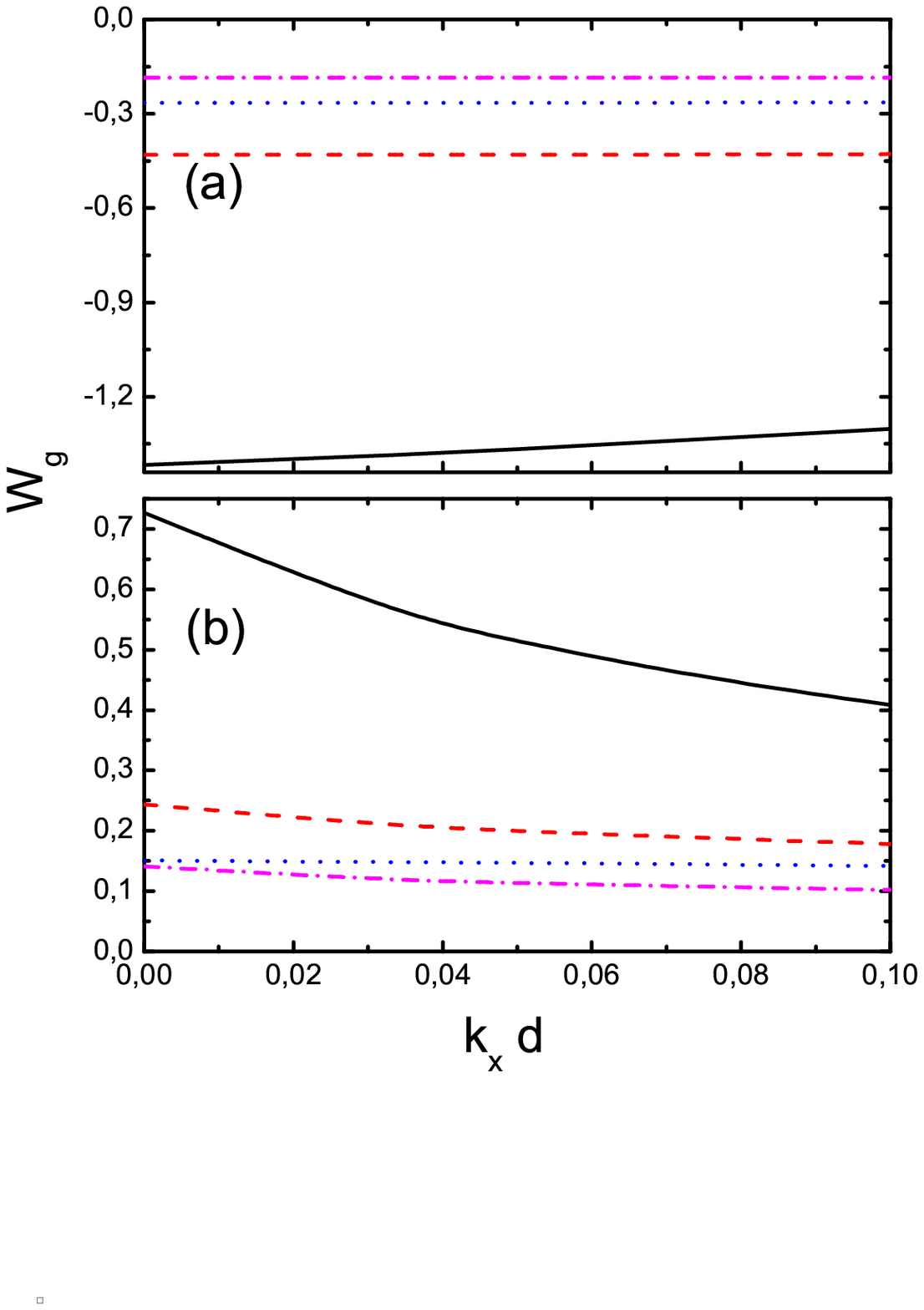}
\vspace*{-3.0cm}
\caption{(Color online) Dimensionless group velocity $W_{g}$ as function of $k_{x}d$ at a long-wavelength region
$0.1 \geq k_{x} d > 0$ for the eight waves of Fig. 2; $\eta=1$, $2M_{0}/B_{ext}=0.5$, $B_{ext}=0.6$T. }
\end{figure}

\begin{figure}[ht]
\vspace*{-0.3cm}
\includegraphics [height=14cm, width=10cm]{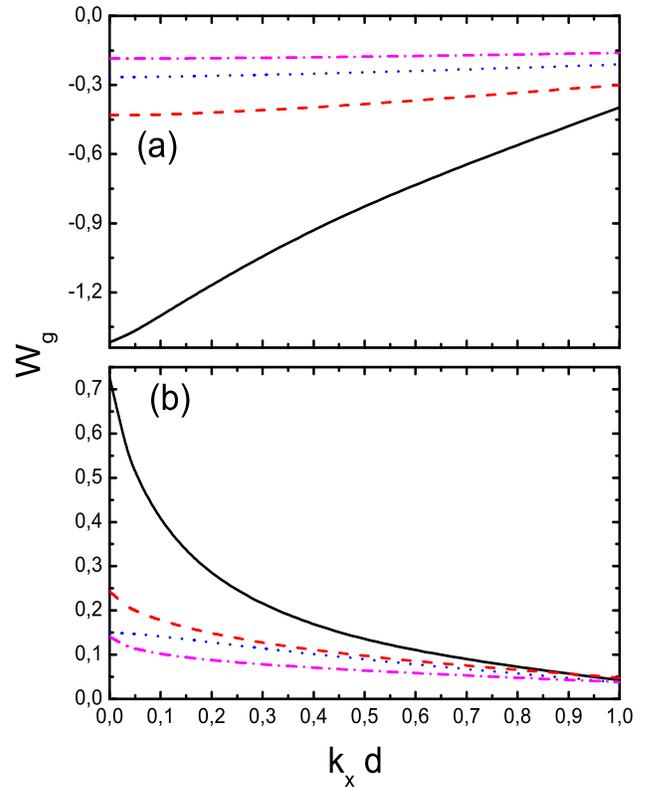}
\vspace*{-3.0cm}
\caption{(Color online) Dimensionless group velocity $W_{g}$ as function of $k_{x}d$ at 
$1.0 \geq k_{x} d > 0$ for parameters of Fig. 2.
Fig. 5(a) and Fig. 5(b) show 
MEMPs with the negative and the positive chirality.}
\end{figure}

\subsection{Magnetic edge magnetoplasmons for $B_{ext} >0$, $M_{0}>0$ and $B_{ext} < 0$, $M_{0}<0$}

\subsubsection{Magnetic edge magnetoplasmons for $B_{ext} >0$, $M_{0}>0$}

In Fig. 2(a) we present the dispersion relations for dimensionless phase velocities, $W=W_{j}(k_{x})$,  
of four fastest MEMP modes $|W^{(n)}_{1}|>|W^{(n)}_{2}|>...>|W^{(n)}_{4}|$ 
with the negative chirality, $W<0$, and of four fastest MEMP modes
with the positive chirality, $W^{(p)}_{1}>W^{(p)}_{2}>...>W^{(p)}_{4} >0$.  
Fig. 2 is plotted for $\eta=1.0$, and $2M_{0}/B_{ext}=0.5$; the latter
for $B_{ext}=0.6$T implies that $M_{0}=B^{2}_{ext}/(4B_{0})=3/20$T for $B_{0}=0.6$T.
Fig. 2(b) presents a zoom of the dispersion relation $W^{(n)}_{1}(k_{x}) \approx 0.53 \, k_{x}d-1.417$ and it is seen that
$W^{(n)}_{1}(k_{x}) $  very close follow the  linear dependence,  on $k_{x}$; according to Fig. 2(a), the rest three MEMPs 
of the negative chirality have negligible spatial dispersion in present long-wavelength region as
their $W^{(n)}_{j}(k_{x})$ , $j=2, 3, 4$,   are practically independent of $k_{x}$.
Fig. 2(c) presents a zoom of the dispersion relation $W^{(p)}_{1}(k_{x}) \approx -2.03 \, k_{x}d+0.727$ which rather closely follow
the linear dependence on  $k_{x}$; in addition, it is much steeper than in Fig. 2(b).
The rest of MEMPs with the positive 
chirality $W^{(p)}_{j}(k_{x})$, $j=2, 3, 4$, in Fig. 2(a) have an essential spatial dispersion of qualitatively  the same form as
$W^{(p)}_{1}(k_{x})$, however, with much smaller dependence on $k_{x}$.
Results for MEMPs presented  in Fig. 2 (and other figures of the work) correspond to a very good convergence
for used $N_{0}=11$; here from the $24 \times 24$ matrix system of equations we obtain 24 MEMPs. 

\begin{figure}[ht]
\vspace*{-0.3cm}
\includegraphics [height=14cm, width=10cm]{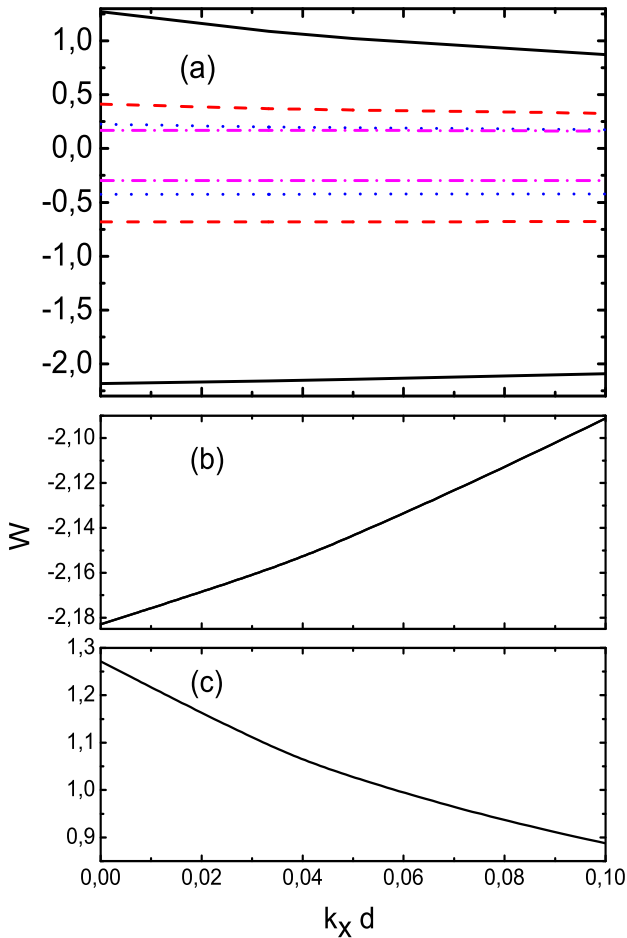}
\vspace*{-1.5cm}
\caption{(Color online) Dimensionless phase velocity $W$ as function of $k_{x}d$ 
at a long-wavelength region $0.1 \geq k_{x} d > 0$ for
$\eta=2$, $2M_{0}/B_{ext}=0.5$, $B_{ext}=0.6$T. Figs. 6(a), 6(b), and 6(c) show: 
(a) four fastest MEMPs among the modes with the positive (negative) chirality,
(b) zoom for the fastest MEMP with the negative chirality, and (c) zoom for the fastest MEMP with the positive chirality.}
\end{figure}

\begin{figure}[ht]
\vspace*{-0.3cm}
\includegraphics [height=14cm, width=10cm]{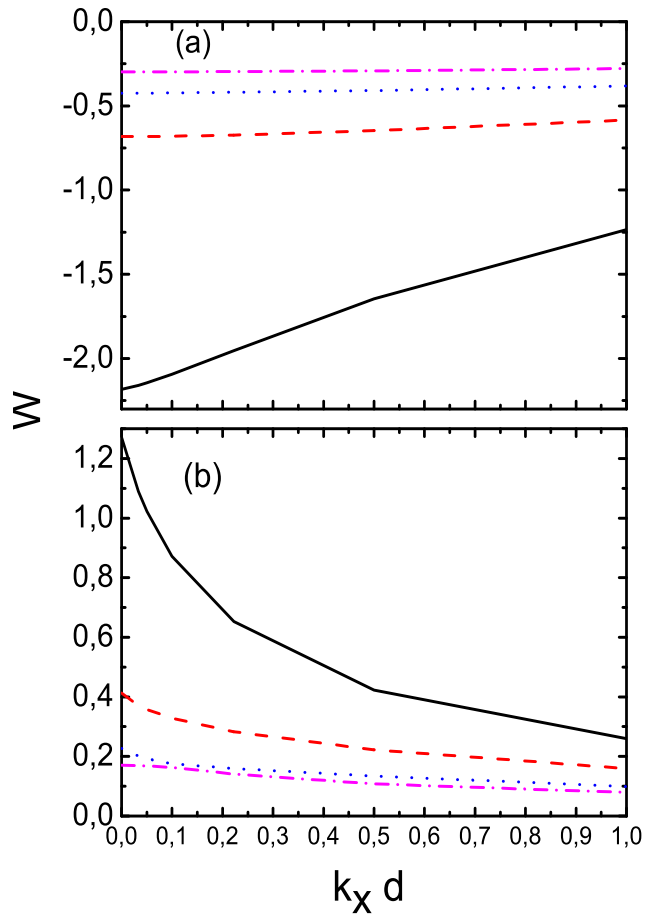}
\vspace*{-2.0cm}
\caption{(Color online) Dimensionless phase velocity $W$ as function of $k_{x}d$ at a wide region
$1.0 \geq k_{x} d > 0$ for parameters of Fig. 6.
Figs. 7(a) and 7(b) show 
four fastest MEMPs among the modes with the negative  and the positive chirality, respectively.}
\end{figure}

Point out that here without the ferromagnetic film (with the same direction of $\mathbf{B}_{ext}$)
for a usual type of the edge at $y=0$, 
due to a  monotonic decrease of the electron density for increasing $y$, it follows \cite{aleiner94,balev00}
that $W<0$ for all EMPs; i.e., the EMPs have the same negative chirality as present $W^{(n)}_{i}$  MEMPs. 

In Fig. 3a and Fig. 3b we present the dispersion relations, $W=W_{j}(k_{x})$,  for eight MEMPs
shown in Fig. 2a at a wider region  $1.0 \geq k_{x} d > 0$.
It is seen in Fig. 3b that $W^{(p)}_{1}(k_{x})$ for $1 \geq k_{x} d \gg 0.1$ essentially changes its dependence 
on $k_{x}$ from a linear one to approximately $\propto k_{x}^{-1}$, however, in Fig. 3a the changes
for $W^{(n)}_{1}(k_{x})$ are rather small. In addition, Fig 3a shows that only at 
$k_{x} d \sim 1$ velocities $W^{(n)}_{j}(k_{x})$ , $j=2, 3, 4$,  become a weakly dependent on $k_{x}$.

\begin{figure}[ht]
\vspace*{-0.3cm}
\includegraphics [height=14cm, width=9.5cm]{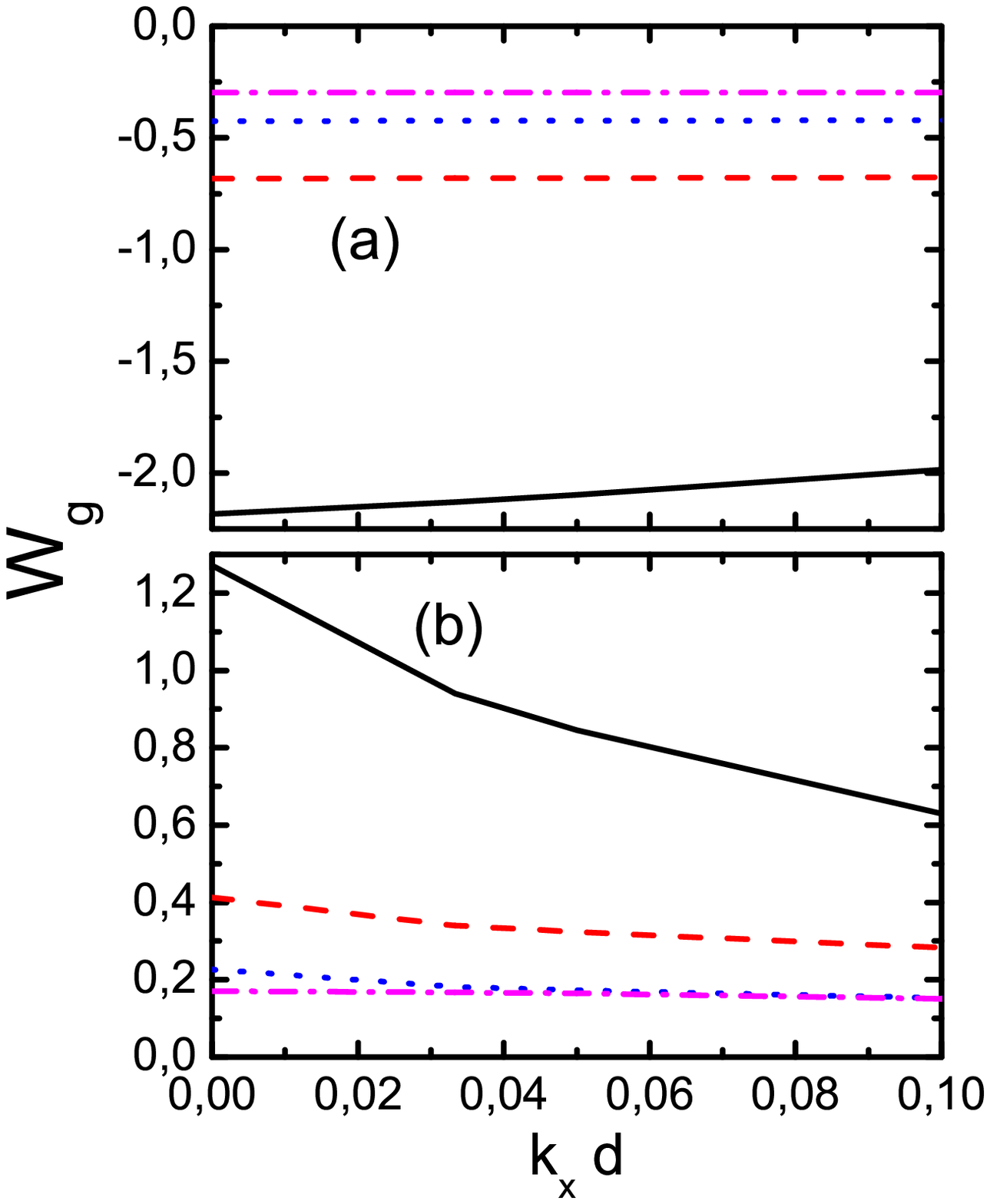}
\vspace*{-1.5cm}
\caption{(Color online) Dimensionless group velocity $W_{g}$ as function of $k_{x}d$ at a long-wavelength region
$0.1 \geq k_{x} d > 0$ for the eight waves of Fig. 6 
}
\end{figure}

\begin{figure}[ht]
\vspace*{-0.3cm}
\includegraphics [height=14cm, width=9.5cm]{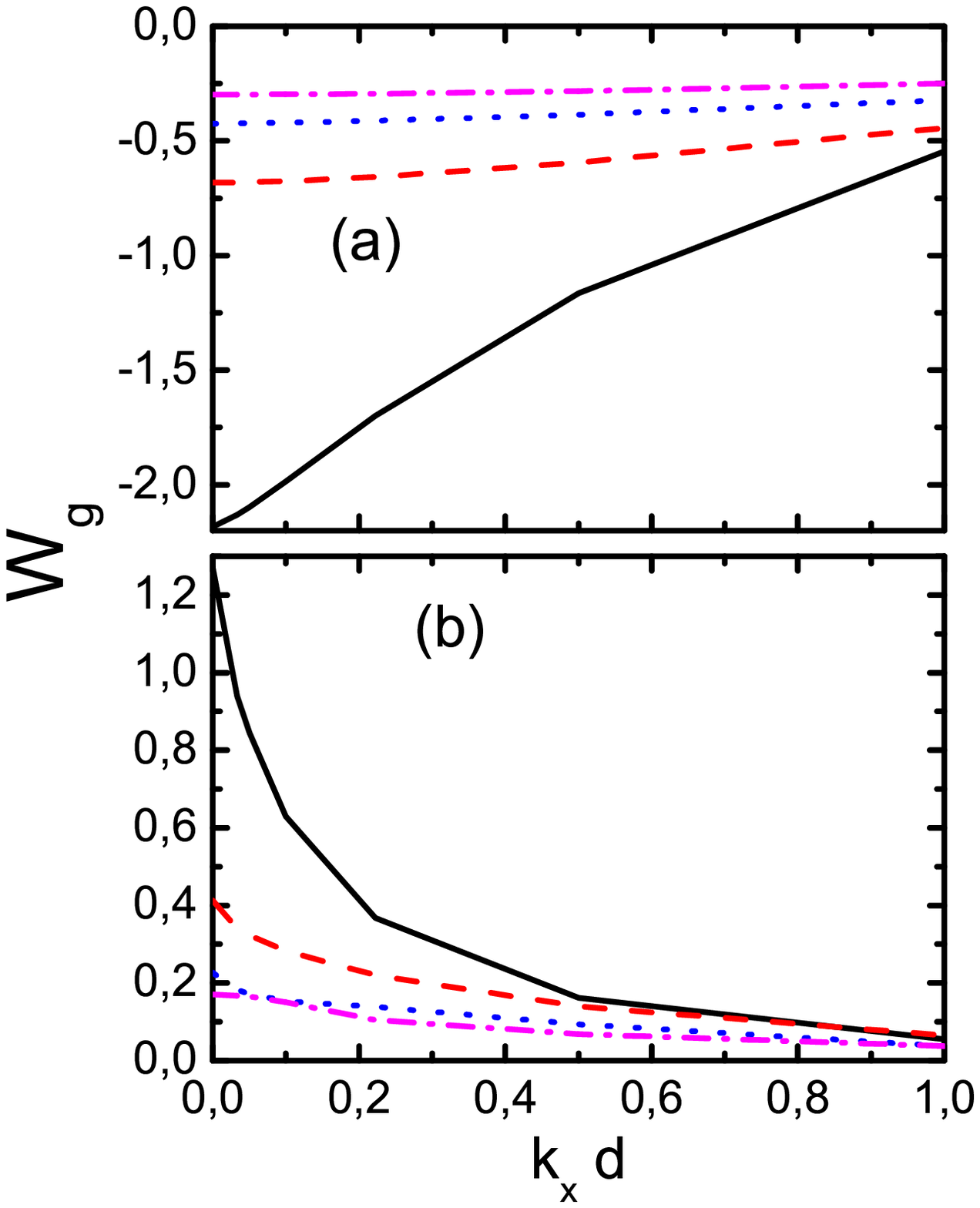}
\vspace*{-1.5cm}
\caption{(Color online) Dimensionless group velocity $W_{g}$ as function of $k_{x}d$ at 
$1.0 \geq k_{x} d > 0$ for parameters of Fig. 6.
Fig. 9(a) and Fig. 9(b) show 
MEMPs with the negative and the positive chirality.}
\end{figure}

\begin{figure}[ht]
\vspace*{-0.3cm}
\includegraphics [height=14cm, width=10cm]{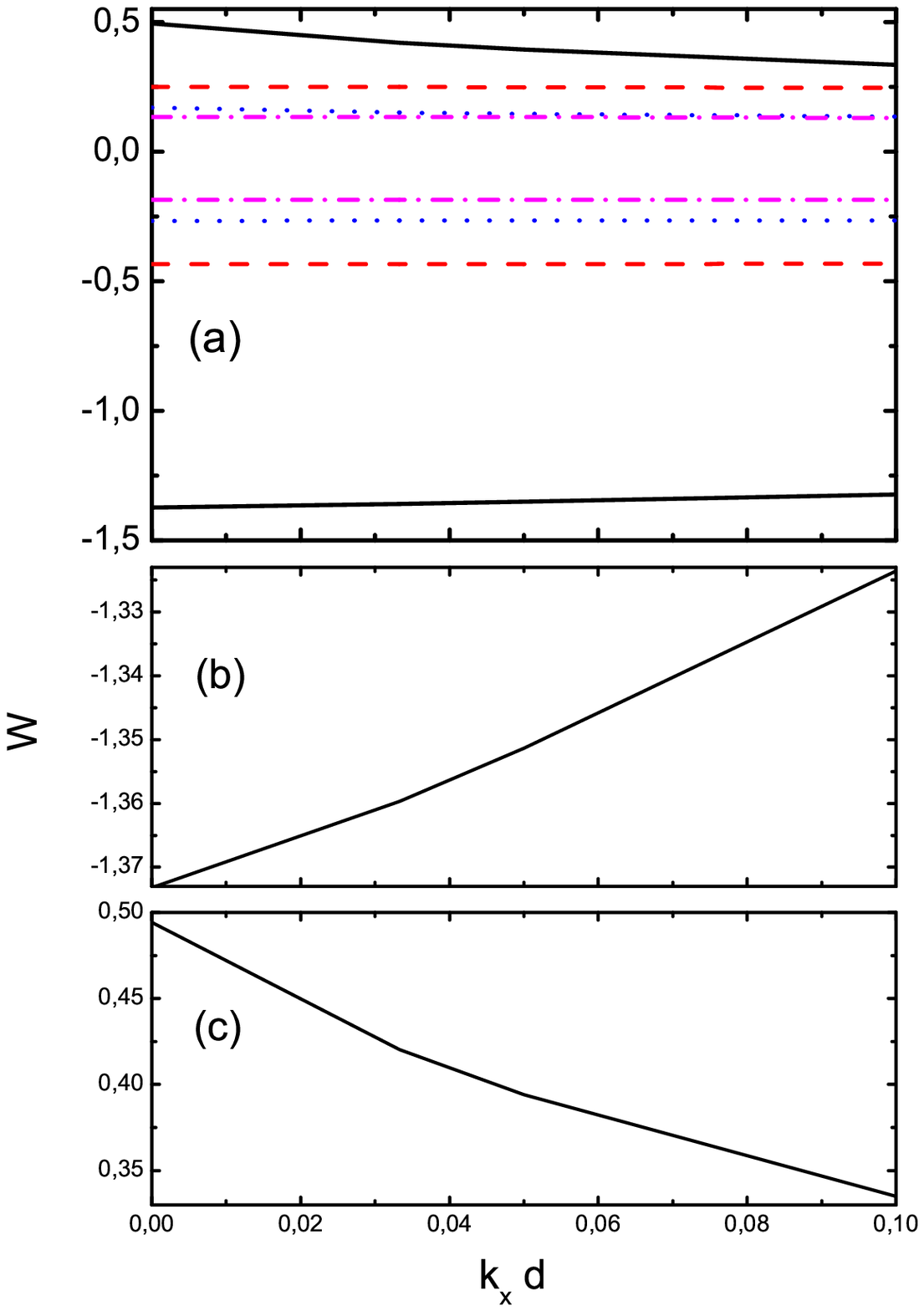}
\vspace*{-1.0cm}
\caption{(Color online) Dimensionless phase velocity $W=\omega/k_{x} v_{0}$ as function of $k_{x}d$ 
at a long-wavelength region $0.1 \geq k_{x} d > 0$ for
$\eta=1$, $2M_{0}/B_{ext}= - 0.5$, $B_{ext}= - 0.6$T, $v_{0} > 0$. Figs. 10(a), 10(b), and 10(c) show: 
(a) four fastest MEMPs among the modes with the positive (negative) chirality,
(b) zoom for the fastest MEMP with the negative chirality, and (c) zoom for the fastest MEMP with the positive chirality.}
\end{figure}

\begin{figure}[ht]
\vspace*{-0.3cm}
\includegraphics [height=14cm, width=10cm]{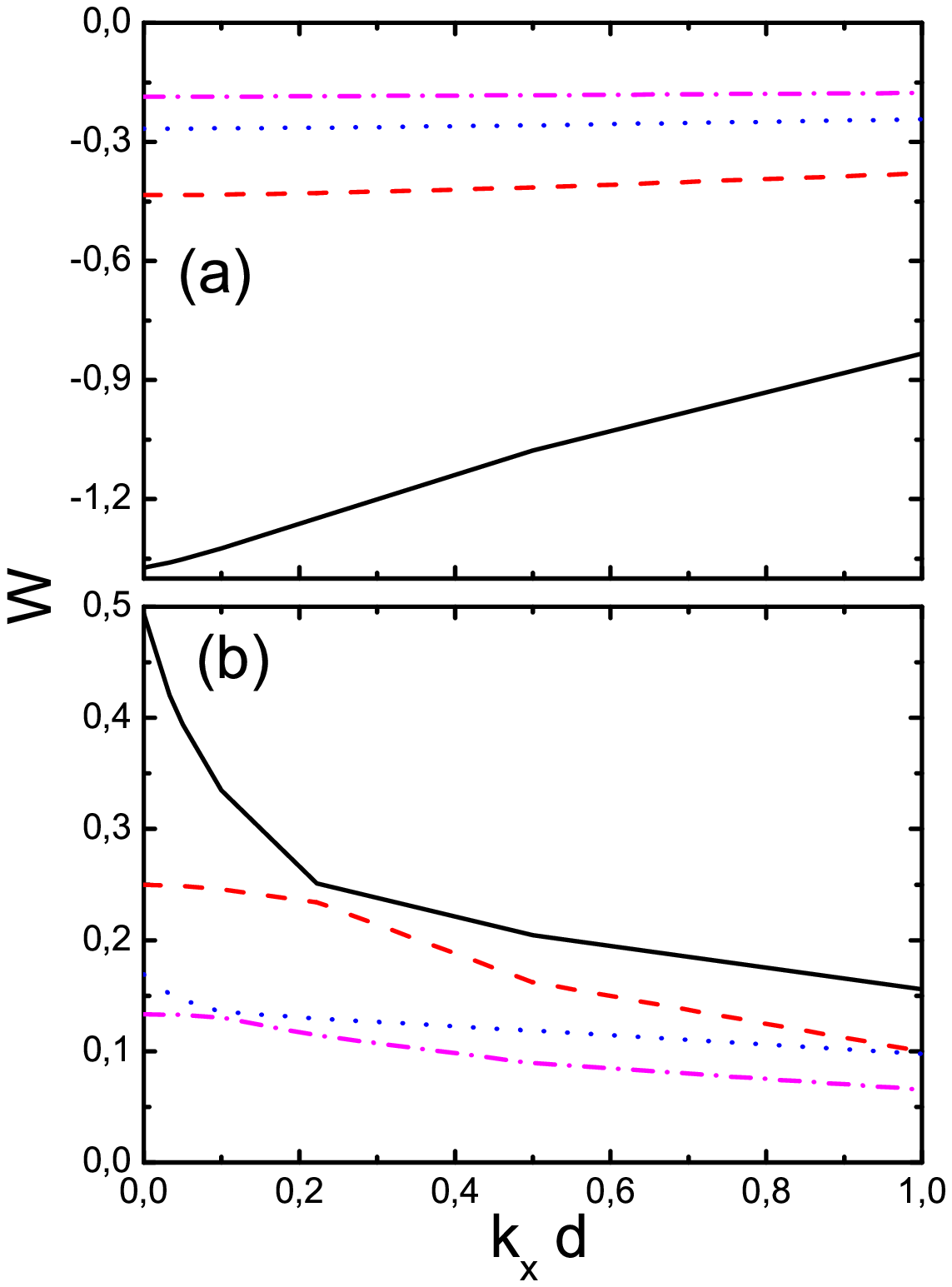}
\vspace*{-2.0cm}
\caption{(Color online) Dimensionless phase velocity $W$ as function of $k_{x}d$ at a wide region
$1.0 \geq k_{x} d > 0$ for parameters of Fig. 10.
Figs. 11(a) and 11(b) show 
four fastest MEMPs among the modes with the negative  and the positive chirality, respectively.
In Fig. 11(b)  three anti-crossings are seen. }
\end{figure}

In Fig. 4(a) and Fig. 4(b) we present dimensionless group velocity, $W_{g}(k_{x})=d \; \omega_{j}(k_{x})/(v_{0}dk_{x})$,  for MEMPs
shown in Fig. 3(a) and Fig. 3(b), at the long-wavelength region  $0.1 \geq k_{x} d > 0$.
In addition, in Fig. 5(a) and Fig. 5(b) we present dimensionless group velocity, $W_{g}$,  
for MEMPs shown in Fig. 3(a) and Fig. 3(b), at a wide region  $1.0 \geq k_{x} d > 0$.

In Fig. 6 for $\eta=2$ we plot the same dependences as in Fig. 2 for parameters that coincide, except of $\eta$, with
those of Fig. 2.
In Fig. 6(a) we present the dispersion relations for the dimensionless phase velocities, $W=W_{j}(k_{x})$,  
of four fastest MEMP modes $|W^{(n)}_{1}|>|W^{(n)}_{2}|>...>|W^{(n)}_{4}|$ 
of the negative chirality and of four fastest MEMP modes
of the positive chirality, $W^{(p)}_{1}>W^{(p)}_{2}>...>W^{(p)}_{4}>0$.  
Fig. 6(b) and Fig. 6(c) present a zoom of the dispersion relation 
$W^{(n)}_{1}(k_{x}) \approx 0.89 \, k_{x}d-2.183$ and 
$W^{(p)}_{1}(k_{x}) \approx -4.00 \, k_{x}d+1.271$, respectively.

In Fig. 7(a) and Fig. 7(b) we present the dispersion relations, $W=W_{j}(k_{x})$,  for eight MEMPs
shown in Fig. 6(a) at a wider region  $1.0 \geq k_{x} d > 0$.
It is seen in Fig. 7(b) that $W^{(p)}_{1}(k_{x})$ for $1 \geq k_{x} d \gg 0.1$ essentially changes its dependence 
on $k_{x}$ from the one for $k_{x} d \ll 1$. However, in Fig. 7(a) the changes
for $W^{(n)}_{1}(k_{x})$ are rather small so it is still well approximated as $W^{(n)}_{1}(k_{x}) \approx 0.89 \, k_{x}d-2.183$
in a wide region  $1.0 \geq k_{x} d > 0$. In addition, Fig. 7(a) shows that only at 
$k_{x} d \sim 1$ velocities $W^{(n)}_{j}(k_{x})$ , $j=2, 3, 4$,  become a weakly dependent on $k_{x}$.

In Fig. 8(a) and Fig. 8(b) we present dimensionless group velocity, $W_{g}(k_{x})$,  for MEMPs
shown in Fig. 7(a) and Fig. 7(b), at the long-wavelength region  $0.1 \geq k_{x} d > 0$.
In addition, in Fig. 9(a) and Fig. 9(b) we present $W_{g}$,
for MEMPs shown in Fig. 7(a) and Fig. 7(b), at a wide region  $1.0 \geq k_{x} d > 0$.

\subsubsection{Magnetic edge magnetoplasmons for $B_{ext} < 0$, $M_{0}<0$}

In Figs. 2 - 9 we have studied MEMPs for $B_{ext}>0$, $M_{0} >0$ which give $v_{0} >0$ and $B_{0}>0$
for all these figures. If we change the sign of $B_{ext}$ and $M_{0}$ on negative then 
$v_{0}$ and $B_{0}$ also become negative. Further, it is seen that we again arrive to the same Figs. 2 - 9, however,
now $W=\omega/v_{0} k_{x}>0$ due to $v_{0}<0$ will correspond to the negative chirality (as here $\omega/k_{x} <0$)
and $W<0$ will correspond to the positive chirality (as here $\omega/k_{x} >0$).

\subsection{Effect of a hysteresis. Magnetic edge magnetoplasmons for $B_{ext} <0$, $M_{0}>0$ and $B_{ext} > 0$, $M_{0}<0$.}

Fig. 10 is plotted for the same positive $M_{0}$, $\eta=1.0$, and $|B_{ext}|$ as Fig. 2, however, here $B_{ext} <0$  and,
respectively, $2M_{0}/B_{ext}= - 0.5$.
For $B_{ext}=-0.6$T this implies that $M_{0}=B^{2}_{ext}/(4B_{0})=3/20$T for $B_{0}=0.6$T.
Point out that $v_{0}$ and $B_{0}$ in Fig. 10 have the same positive values as in Fig. 2.
In Fig. 10(a) we present the dispersion relations for dimensionless phase velocities, $W=W_{j}(k_{x})$,  
of four fastest MEMP modes $|W^{(n)}_{1}|>|W^{(n)}_{2}|>...>|W^{(n)}_{4}|$ 
with the negative chirality, $W<0$, and of four fastest MEMP modes
with the positive chirality, $W^{(p)}_{1}>W^{(p)}_{2}>...>W^{(p)}_{4} >0$ at a long-wavelength region  $k_{x} d \ll 1$.
Fig. 10(b) presents a zoom of the dispersion relation $W^{(n)}_{1}(k_{x}) \approx 0.50 \, k_{x}d-1.373$. 
According to Fig. 10(a), the rest MEMPs 
of the negative chirality have negligible spatial dispersion in the long-wavelength region.
Fig. 10(c) presents a zoom of the dispersion relation $W^{(p)}_{1}(k_{x}) \approx -1.59 \, k_{x}d+0.494$ ; 
it is much steeper than in Fig. 10(b).
From comparison of Fig. 10(b) and Fig. 10(c) with Figs. 2(b), 2(c) it is seen that the velocity of the fastest MEMP with the negative
chirality ($W<0$)  only weakly decreases, about 3\%, for a reversed sign of $B_{ext}$ as the velocity of the fastest
MEMP with the positive chirality decreases much stronger, more than 30\%.

In Fig. 11 dimensionless phase velocity $W$ is plotted for a wider region of $k_{x}d$ than in Fig. 10; parameters
are the same as in Fig. 10.  In Fig. 11(b) two anti-crossings for phase velocities are clearly seen for $k_{x} d < 0.3$: one at $k_{x}d \approx 0.1$ and
another at $k_{x}d \approx 0.2$.

In Figs. 10 - 11 we have studied MEMPs for $B_{ext}<0$, $M_{0} >0$ which give $v_{0} >0$, $B_{0}>0$, and 
$2M_{0}/B_{ext}= - 0.5$ for these figures. If we change the sign of $B_{ext}$ on positive and and the sign of $M_{0}$ on negative then 
both $v_{0}$ and $B_{0}$ become negative. Let us assume that the absolute values of parameters are the same as in Figs. 10 - 11.
Then it is seen that we will obtain the same Figs. 10 - 11, however,
where now $W=\omega/v_{0} k_{x}>0$ due to $v_{0}<0$ will correspond to the negative chirality ($\omega/k_{x} <0$)
and $W<0$ will correspond to the positive chirality ($\omega/k_{x} >0$).

\section{Concluding Remarks} 

We have shown that HPMG strongly changes properties of MEMPs. In particular,
they obtained spatial dispersion of unconventional form in the long-wavelength region, $k_{x}d \leq 0.1$. 
Where for two most fast  MEMPs the phase 
velocities $\omega_{1}^{(n,p)}(k_{x})/k_{x }\approx  (a_{n,p} k_{x}+b_{n,p})$.
In a wider region $1 \geq k_{x} d \geq 0$
the phase velocity of most fast MEMP (with the negative chirality, $W^{(n)} <0$, $b_{n} <0$) 
still rather closely follow the same linear behaviour, $\omega_{1}^{(n)}(k_{x})/k_{x }\approx  (a_{n} k_{x}+b_{n})$,
e.g., see Figs. 3(a), 7(a), 11(a).
However,  the phase velocity of second most fast MEMP 
(with the positive chirality, $W^{(p)}>0$, $b_{p} > 0$)  for $1 \geq k_{x}d \geq 0.1$ shows behaviour of 
$\omega_{1}^{(p)}(k_{x})/k_{x }$ essentially different from $\approx  (a_{p} k_{x}+b_{p})$,
e.g., see Figs. 3(b), 7(b), 11(b).

Point out, two most fast MEMPs $\omega_{1}^{(n,p)}(k_{x})$ propagate in opposite 
directions along the magnetic edge, with a strong overlap and localization of their spatial structures  
at the edge region. Then appearance of essential imperfections (e.g., they can be created on purpose)
at the ends of a segment of length $\Lambda_{x}$ along the magnetic edge
will  create a coupling at these ends between both MEMPs and at some resonance frequency $\omega_{r}$ a closed wave path 
within the section $\Lambda_{x}$. If the phase shift due to reflection at the ends of the section is negligible and 
the wave vectors $k_{x}^{(n),(p)}>0$ are so small that
$\omega_{r} \approx |b_{n}| k_{x}^{(n)} = b_{p} k_{x}^{(p)}$ then from
 $(k_{x}^{(n)}+k_{x}^{(p)}) \Lambda_{x}=2\pi N$, with $N=1, 2,...$,
 we obtain that
 \begin{equation}
 k_{x}^{(n)} =\frac{b_{p}}{|b_{n}|} k_{x}^{(p)}= \frac{2\pi N b_{p}}{\Lambda_{x} (|b_{n}|+b_{p})} .
 \label{eq16}
\end{equation}%

In our model, see Fig. 1,  we assume that  HPMG is given by  the half-plane   
metallic nonmagnetic film of negligible thickness, ($y<0$, $z=d$), 
and that 2DES is embedded in GaAs based sample.
Notice, for a real experimental setup a finite thickness of HPMG   can be treated as negligible 
if it is much smaller than $d$, and $\eta d$. 
Besides HPMG 
and the plane of 2DES the rest of space we assume as a dielectric background with  
spatially homogeneous (and real) dielectric constant $\varepsilon$; i.e.,   
the ferromagnetic semi-infinite film is a dielectric one with dielectric constant  $\varepsilon \approx 12.5$.
Point out, such ferromagnetic materials as ferrite or ceramic magnets can show both dielectric
and ferromagnetic properties similar to ones assumed in present study.

In addition, the properties of MEMPs for the present model, at $\eta \leq 1$ (especially for $\eta \ll 1$), should be similar 
to the properties of MEMPs of more easily experimentally realised
setup with the ferromagnetic semi-infinite film, as in Fig. 1, that is also  \textit{metallic} and placed on a top of GaAs sample
(with the rest of a space filled in, e.g., with the dielectric medium with the same dielectric constant as GaAs, except the $z=0$
plane of 2DES). In this case the ferromagnetic semi-infinite film can be made, e.g., from Dysprosium (Dy) or Permalloy that
at liquid helium temperature $T=4.2$K will show necessary ferromagnetic properties and, due to their metallic behaviour, 
well approximate the effect of HPMG. The decrease of $\eta$ between $1$ to $0.1$ will
 mainly lead, e.g., to a decrease of $|W_{1}^{(n),(p)}(k_{x}=0)| \propto \eta$; so that for $\eta=0.1$ (other parameters
are the same as in Fig. 2)  it follows $W_{1}^{(n)}(k_{x})=0.07 k_{x}d-0.210$ and $W_{1}^{(p)}(k_{x})=0.085-0.20 k_{x}d$.
This decrease of $\eta$ has minor effect on the spatial structure of MEMPs or a form of lateral inhomogeneity of the magnetic
field at the  2DES plane; however, it essentially decreases the characteristic amplitude of the latter modulation;
the latter decrease becomes $\propto \eta$. for $\eta^{2} \ll 1$. Point out for 2DES with the density 
$n_{I} \approx 2 \times 10^{11} cm^{-2}$ in GaAs based heterostructure at $T=4.2$K  usually
the mobility $u=|e| \tau/m^{\ast} \approx 10^{6} cm^{2}/V s$. Then the condition of the strong magnetic 
field $ \omega_{c}(y) \tau \gg 1$ gives $|B(y)| \gg 0.01$T; in addition, from Eq. (\ref{eq1}),  for
$\eta \leq 2$,  it follows that this condition reduces to $|B_{ext}| \gg 0.01$T.

Present MEMPs are slow, $\omega/k_{x} \ll c/\sqrt{\varepsilon}$, potential waves; as $\eta\;v_{0}$ is 
a characteristic phase velocity of
the MEMPs,  this condition obtains the form $B^{2}_{ext} \gg 4\eta |e| n_{I} M_{0}/\sqrt{\varepsilon}$.
In addition, the results reported in this work, for  $k_{x} d \leq 1$, assume that $v_{0} \ll \omega_{c}(y) d$ 
and a strong magnetic moment, 
$2|e|M_{0}/m^{\ast}c \gg \omega \cdot k_{x} d$.

\begin{acknowledgments}
This work was supported by the Brazilian FAPEAM  (Funda\c{c}\~{a}o de Amparo \`{a} Pesquisa do Estado do
Amazonas) Grants: Universal Amazonas (Edital 021/2011), O. G. B., and PVS (Pesquisador Visitante Senior), I. A. L.
Also support, I. A. L., by  funding from
the European Union Seventh Framework Programme (FP7/2007-2013) under grant
agreement n$^{\mathrm{o}}$~PCOFUND-GA-2009-246542 and the Foundation for Science and Technology of Portugal
is acknowledged.
\end{acknowledgments}
\appendix
\section{Matrix elements}\label{app1}

\begin{eqnarray}
&&I^{(1)}_{mn}=\frac{\pi }{2}\int_{0}^{1}dX\cos (m \pi
X) \int_{0}^{1}dX^{\prime }  \left[ \cos (n \pi X^{\prime })-\frac{1}{2}\delta _{n,0}\right] \nonumber \\ 
&&\times \left[ \frac{1+\eta }{1+\eta (2+\eta )\cos^{2}(\pi X^{\prime }/2)}-1\right]    
\left\{  f_{0}(X^{\prime })  \left[ \Delta_{\varphi}(k_{x},X,X^{\prime})  \right.   \right.  \nonumber \\
&&\left. +\Delta_{\varphi}(k_{x},-X,X^{\prime})    \right]+ f_{0}(-X^{\prime }) \nonumber \\
&&\times \left.  \left[ \Delta_{\varphi}(k_{x},X,-X^{\prime})+   
\Delta_{\varphi}(k_{x},-X,-X^{\prime})    \right] \right\} ,  
\label{eqa1}
\end{eqnarray}%
where $m \leq N_{0}$, $n \leq N_{0}$;
\begin{eqnarray}
&&I^{(2)}_{mn}=\frac{\pi }{2}\int_{0}^{1}dX\sin \left[ (m-N_{0}) \pi X \right] 
\int_{0}^{1}dX^{\prime }  \left[ \cos (n \pi X^{\prime })  \right. \nonumber \\ 
&& \left.  -\frac{1}{2}\delta _{n,0}\right]  \left[ \frac{1+\eta }{1+\eta (2+\eta )\cos^{2}(\pi X^{\prime }/2)}-1\right]    
\left\{  f_{0}(X^{\prime })     \right.  \nonumber \\
&&\times \left[ \Delta_{\varphi}(k_{x},X,X^{\prime})  
 -\Delta_{\varphi}(k_{x},-X,X^{\prime})    \right]+ f_{0}(-X^{\prime }) \nonumber \\
&&\times \left.  \left[ \Delta_{\varphi}(k_{x},X,-X^{\prime})-   
\Delta_{\varphi}(k_{x},-X,-X^{\prime})    \right] \right\} ,  
\label{eqa2}
\end{eqnarray}%
where $m \geq N_{0}+1$, $n \leq N_{0}$;
\begin{eqnarray}
&&I^{(3)}_{mn}=\frac{\pi }{2}\int_{0}^{1}dX\cos (m \pi
X) \int_{0}^{1}dX^{\prime }   \sin \left[ (n-N_{0}) \pi X^{\prime } \right]  \nonumber \\ 
&&\times \left[ \frac{1+\eta }{1+\eta (2+\eta )\cos^{2}(\pi X^{\prime }/2)}-1\right]    
\left\{  f_{0}(X^{\prime })  \left[ \Delta_{\varphi}(k_{x},X,X^{\prime})  \right.   \right.  \nonumber \\
&&\left. +\Delta_{\varphi}(k_{x},-X,X^{\prime})    \right]- f_{0}(-X^{\prime }) \nonumber \\
&&\times \left.  \left[ \Delta_{\varphi}(k_{x},X,-X^{\prime})+   
\Delta_{\varphi}(k_{x},-X,-X^{\prime})    \right] \right\} ,  
\label{eqa3}
\end{eqnarray}%
where $m \leq N_{0}$, $n \geq N_{0}+1$;
\begin{eqnarray}
&&I^{(4)}_{mn}=\frac{\pi }{2}\int_{0}^{1}dX  \sin \left[ (m-N_{0}) \pi X \right] 
 \int_{0}^{1}dX^{\prime }   \sin \left[ (n-N_{0}) \pi X^{\prime } \right]  \nonumber \\ 
&&\times \left[ \frac{1+\eta }{1+\eta (2+\eta )\cos^{2}(\pi X^{\prime }/2)}-1\right]    
\left\{  f_{0}(X^{\prime })  \left[ \Delta_{\varphi}(k_{x},X,X^{\prime})  \right.   \right.  \nonumber \\
&&\left. -\Delta_{\varphi}(k_{x},-X,X^{\prime})    \right]- f_{0}(-X^{\prime }) \nonumber \\
&&\times \left.  \left[ \Delta_{\varphi}(k_{x},X,-X^{\prime})-   
\Delta_{\varphi}(k_{x},-X,-X^{\prime})    \right] \right\} , 
\label{eqa4}
\end{eqnarray}%
where $m \geq N_{0}+1$, $n \geq N_{0}+1$.

\end{document}